\documentstyle[aps,preprint,prc]{revtex}
\tighten

\input{boxedeps.tex}
\SetRokickiEPSFSpecial
\HideDisplacementBoxes

\begin{document}
\draft

\title{Equivalent Photon Approach to Simultaneous Excitation in Heavy
Ion Collision}

\author{
Kai Hencken\footnote{EMail address: hencken\@phys.washington.edu}
}
\address{
Institut f\"ur theoretische Physik der Universit\"at Basel,
Klingelbergstrasse 82, 4056 Basel, Switzerland
}
\address{
Institute for Nuclear Theory, University of Washington,
Physics/Astronomy Building, Box 351550, Seattle, WA 98195, USA
\footnote{present address}
}

\author{
Dirk Trautmann
}
\address{
Institut f\"ur theoretische Physik der Universit\"at Basel,
Klingelbergstrasse 82, 4056 Basel, Switzerland
}
\author{
Gerhard Baur
}
\address{
Institut f\"ur Kernphysik (Theorie), Forschungszentrum J\"ulich,
52425 J\"ulich, Germany
}

\date{\today}

\maketitle

\begin{abstract}
We apply the Equivalent Photon Approximation to calculate cross
sections for the simultaneous excitation of two heavy ions in
relativistic collisions. We study especially the excitation of two
nuclei to a $1^-$-state and show that the equations are symmetric
with respect to both ions. We also examine the limit in which the
excitation energy of one of the nuclei goes to zero, which gives the
elastic case.  Finally a few remarks about the limits of this
approach are made.
\end{abstract}


\narrowtext

\section{Introduction}
\label{Sec:Intro}

Recently \cite{hencken95} a formalism was presented that allows the
calculation of the equivalent photon spectra corresponding to nuclear
transitions of relativistic nuclei ($\gamma \gg 1$). In this way it
became possible to study corrections such as the influence of nuclear
excitations on two-photon processes.  The example that was studied
was one nucleus making a transition to the Giant Dipole Resonance
state (GDR), while at the same time emitting an equivalent photon. It
is the purpose of this brief report to study simultaneous projectile
and target excitation using the equivalent photon method. Using this
method, we will see that the results derived previously by Benesh and
Friar \cite{benesh94} can be simplified very much and the physics
behind this process is made more transparent.

Throughout this report we assume that we can always use the long
wavelength approximation. This requires that $kR < 1$, where $k$ is
the space part of the photon momentum and $R$ the radius of the
nucleus.  From this we conclude, that the excitation energy $\Delta$
and therefore also the energy of the photon $\omega$ has to be low,
that is, $\omega \approx \Delta < 1 / R$.

In the equivalent photon approximation (EPA) the cross section for
the simultaneous excitation is expressed as
\begin{equation}
\sigma_{\Delta_1 \Delta_2} = \int \frac{d\omega}{\omega}\;
 n_{\Delta_1}(\omega) \; \sigma_{\gamma\Delta_2}(\omega),
\label{Eq:epa}
\end{equation}
where $n_{\Delta_1}(\omega)$ is the equivalent photon number of the
inelastic photon emission process, and $\sigma_{\gamma\Delta_2}$ is
the cross section for the photoabsorption process for a photon of
energy $\omega$ in the rest frame of the nucleus.

In \cite{hencken95} the equivalent photon number for the inelastic
photon emission was derived neglecting the longitudinal components of
the electromagnetic field. In the limit of high ion energy ($\gamma
\gg 1$) the result is
\begin{equation}
n_{\Delta}(\omega) = \int \frac{-2 \omega^2 C 
  + q_{\perp}^2 M^2 \gamma^2 D}
  {(2\pi)^3 2 M^2 \gamma^2 (-q^2)^2} d^2q_{\perp}.
\label{Eq:niepa}
\end{equation}
Here $q$ is the four-momentum transfer and $C$ and $D$ are given
generally by
\begin{eqnarray}
C &=& - 2 \pi \left[ |T^e|^2 + |T^m|^2 \right] \\
D &=& \frac{(-q^2)^2}{(\Delta^2-q^2)^2 M^2} 2 \pi \nonumber\\
&&\times \left[2 |M^C|^2 +
\frac{\Delta^2-q^2}{-q^2}\left( |T^e|^2 + |T^m|^2 \right)  \right],
\end{eqnarray}
where $M^C$, $T^e$, and $T^m$ are the usual Coulomb, electric, and
magnetic matrix elements and $-q^2$ is determined by the kinematics
of this process as
\begin{equation}
-q^2 = \frac{\omega^2}{\gamma^2} + 2 \frac{\omega\Delta}{\gamma} +
\frac{\Delta^2}{\gamma^2} + q_{\perp}^2
= q_{\text{min}}^2 + q_{\perp}^2.
\label{Eq:qmin}
\end{equation}
For high values of $\gamma$ we can safely neglect the term
proportional to $C$ compared to $D$ in Eq.~(\ref{Eq:niepa}). The
result is
\begin{eqnarray}
n_{\Delta}(\omega) &=& \int \frac{q_{\perp}^2} 
{(2\pi)^2 2 (\Delta^2-q^2)^2 M^2} 
\biggl[ 2 \left| M^C \right|^2 \nonumber\\
&&+ \frac{\Delta^2-q^2}{-q^2}
\left( \left|T^e\right|^2 + \left|T^m\right|^2\right)\biggr] 
d^2q_{\perp}.
\end{eqnarray}

Using the long wavelength limit, we can write the matrix elements in
terms of the $B(EJ)$ and $B(MJ)$ values (see, e. g.,
\cite{shalit74}). Assuming that a single value of $J$ dominates, the
formulas are:
\begin{eqnarray}
\left| M^C\right|^2 &=& 4 M^2 \frac{(\Delta^2-q^2)^{J}}{[(2J+1)!!]^2} 
\alpha B(EJ)
\label{Eq:MC}\\
\left| T^e\right|^2 &=& 4 M^2 \frac{\Delta^2}{\Delta^2-q^2} 
\frac{J+1}{J} \frac{(\Delta^2-q^2)^{J}}{[(2J+1)!!]^2} \alpha B(EJ)
\label{Eq:TE}\\
\left| T^m\right|^2 &=& 4 M^2 \frac{J+1}{J} 
\frac{(\Delta^2-q^2)^{J}}{[(2J+1)!!]^2} \alpha B(MJ).
\label{Eq:TM}
\end{eqnarray}

Similarly the cross section for the photoabsorption can be written as follows
in the narrow resonance limit,
\begin{equation}
\sigma_{\gamma}(\omega) = \frac{\pi^2}{2 \Delta M^2} 
\left(\left| T^e\right|^2 + \left| T^m\right|^2 \right)
\delta(\omega - \Delta).
\end{equation}
For the photoabsorption of on-shell photons we have of course to set 
also $q^2=0$ for the process in
Eqs.~(\ref{Eq:TE}) and~(\ref{Eq:TM}).

\section{Simultaneous excitation to the GDR}
\label{Sec:GDR}

\begin{figure}[tbp]
\begin{center}
\ForceHeight{4cm}
\BoxedEPSF{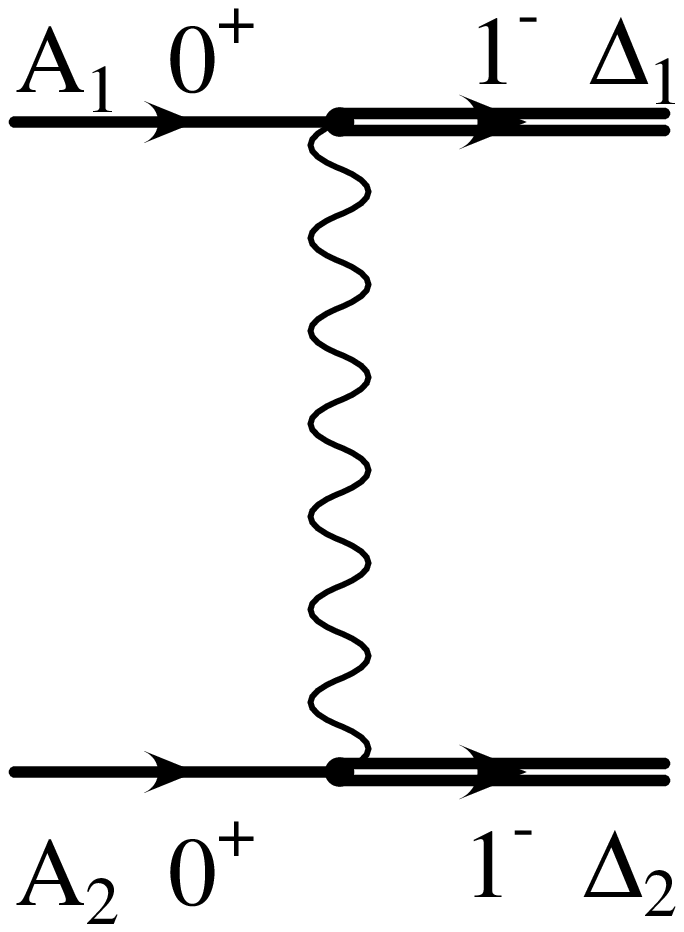}
\end{center}
\caption{Simultaneous excitation of two ions $A_1$ and $A_2$ to the 
giant dipole resonance (GDR). The excitation energy of each GDR is 
denoted by $\Delta_1$ and $\Delta_2$.}
\label{Fig:mutual}
\end{figure}

Let us examine the simultaneous excitation of the GDR of two nuclei
$A_1$ and $A_2$. The corresponding Feynman diagram is shown in
Fig.~\ref{Fig:mutual}. We can view this process in two different
ways, depending on which we consider the emitter and absorber of the
equivalent photon. Let us assume that nucleus $A_2$ is at rest and is
excited to the GDR by absorption of a photon with energy
$\omega=\Delta_2$, where $\Delta_2$ is the excitation energy of the
GDR. This equivalent photon is emitted by the fast moving nucleus
$A_1$, which is excited to its GDR due to this emission.

The excitation to the $1^-$ GDR is an $E1$ transition, proportional
to $B(E1)$.  Using this in Eq.~(\ref{Eq:epa}), we get the cross
section as
\begin{equation}
\sigma_{\Delta_1 \Delta_2} = \frac{4 \alpha^2}{81}
B(E1,\Delta_1) B(E1,\Delta_2)
\int \frac{q_\perp^2 d^2q_\perp}{(-q^2)}.
\end{equation}
We see that the result is symmetric with respect to both ions. 

When integrating over the transverse momenta, we have to keep in mind
that the finite size of the nucleus limits the range of
$q_{\perp}$. The integrand therefore has to be multiplied by the
square of the form factor $F_i^2(q^2)$ of both ions. Using a Gaussian
form factor for both ions $F_i(q^2)=\exp(-|q^2|/\lambda_i^2)$ as a
model for the inelastic form factor, we can evaluate the equivalent
photon number analytically as
\begin{equation}
n_\Delta(\omega) = \frac{\alpha B(E1)}{9 \pi}
\left[ \frac{\lambda^2}{2} 
e^{-2q_{\text{min}}^2/\lambda^2} - q_{\text{min}}^2 
E_1\left(\frac{2 q_{\text{min}}^2}{\lambda^2}\right)\right],
\label{Eq:gauss}
\end{equation}
where $E_1$ is the usual exponential integral \cite{abramowitz65},
$\lambda$ is given by $\lambda = (1/\lambda_1^2 +
1/\lambda_2^2)^{-1/2}$, and $q_{\text{min}} = \Delta_1^2/\gamma^2 + 2
\Delta_1 \omega / \gamma + \omega^2 / \gamma^2$ as defined in
Eq.~(\ref{Eq:qmin}).

An even simpler expression can be found by cutting off the integration 
over $q_{\perp}$ at $1/R$, where $R$ is approximately the sum of the radii of
the ions. Integrating and keeping only the leading
term for large values of $\gamma$ gives the equivalent photon number
\begin{equation}
n(\omega) = \frac{\alpha B(E1,\Delta_1)}{9 \pi R^2}
\qquad\qquad \mbox{(for $\omega < \gamma / R)$},
\end{equation}
and the cross section
\begin{equation}
\sigma_{\Delta_1\Delta_2} = \frac{4 \pi \alpha^2}{81 R^2} 
 B(E1,\Delta_1) B(E1,\Delta_2).
\label{Eq:approx}
\end{equation}
We can see that the dependence of the cross section on $\Delta_1$ and
$\Delta_2$ drops out. It is essentially only proportional to the
product of the corresponding $B(E1)$-values. The limit of large $\gamma$
in Eq.~(\ref{Eq:gauss}) has the form of Eq.~(\ref{Eq:approx}) with
$R^2$ given by $R^2 = 2/\lambda^2 = \frac{1}{3} (\left< r_A^2 \right>
+ \left< r_B^2 \right>)$. But please note that this result is only
valid for a Gaussian form factor.  As discussed in \cite{hencken95}
(see, e.g., Fig.~10 there) $n(\omega)$ is much more sensitive to the
detailed form of the form factor at large $q^2$ than in the elastic
case due to the higher order of $q_\perp$ in the numerator of the
integral. The cross section is also independent of $\gamma$ unlike
the elastic case where the cross section increases with
$\log(\gamma)$ for large gamma. It can be shown easily that
Eq.~(\ref{Eq:approx}) is valid if $\Delta_i \ll \gamma / R$. As
already shown earlier, it is also symmetric with respect to ions
$A_1$ and $A_2$, so the role of both can be interchanged.
Eq.~(\ref{Eq:approx}) has a similar form as found in \cite{benesh94},
but in contrast to that work our result depends on $R$ and therefore
on the size of the nuclei.

In the Goldhaber-Teller model \cite{goldhaber48,deforest66} or the 
sum-rule model \cite{bertulani88} the $B(E1)$ is given by
\begin{equation}
B(E1,0^+ \rightarrow 1^-,\Delta) = \frac{9}{2 m_N \Delta}
\frac{N Z}{A},
\label{Eq:BE1}
\end{equation}
and therefore depends on the excitation energies. This dependence 
on $\Delta_1$ and $\Delta_2$ remains of course in the final formula.

\begin{figure}[tbp]
\begin{center}
\ForceWidth{8.6cm}
\BoxedEPSF{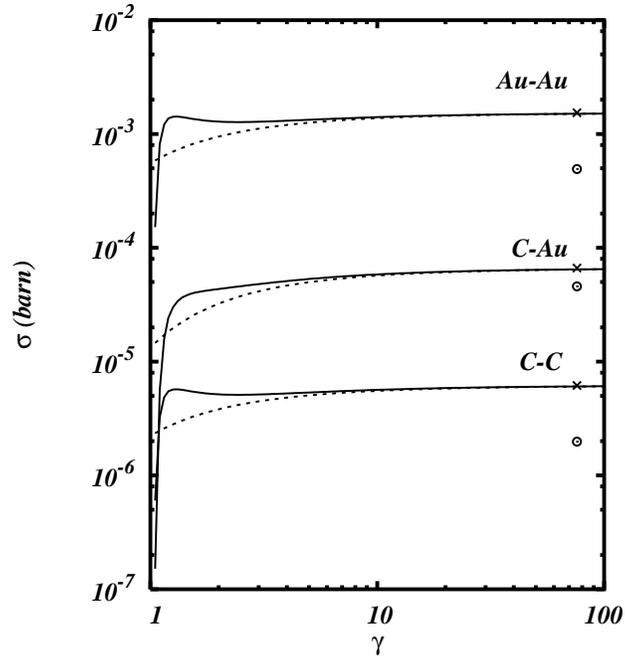}
\end{center}
\caption{Comparison of the different equations for the simultaneous 
excitation cross section as a function of the Lorentz factor
$\gamma$. Shown are the results for $C-C$, $C-Au$, and
$Au-Au$-collisions as examples. The solid line shows the result of the
exact calculation by using the Goldhaber-Teller model and using a
Gaussian form factor, the dashed line the one for the equivalent
photon approximation calculation using also the Gaussian form
factor. The cross is the approximation for large values of $\gamma$
using the cutoff form factor. Also shown are the results of Eq.~(11)
of \protect\cite{benesh94} as circles.}
\label{Fig:cmpsigma}
\end{figure}

In Figure \ref{Fig:cmpsigma} we compare the results of the different
equations.  For the exact calculation we have used the
Goldhaber-Teller model for matrix element and have used a Gaussian
form factor, with $\lambda$ chosen to reproduce the usual $\left< r^2
\right>$ of the nucleus. No further approximation are made.  This is
compared with the results of the equivalent photon approximation of
Eqs.~(\ref{Eq:gauss}) and~(\ref{Eq:approx}). We see that for not too
high energies we get good agreement between all three. Also shown are the
results as given by Eq.~(11) of \cite{benesh94}. They agree reasonable with
our results. Please note that our results depend on the detailed form of the 
form factor at large $q^2$, the good agreement between our three results 
therefore depends also on the use of the same form factor in all of them.

\begin{figure}[tbp]
\begin{center}
\ForceHeight{4cm}
\BoxedEPSF{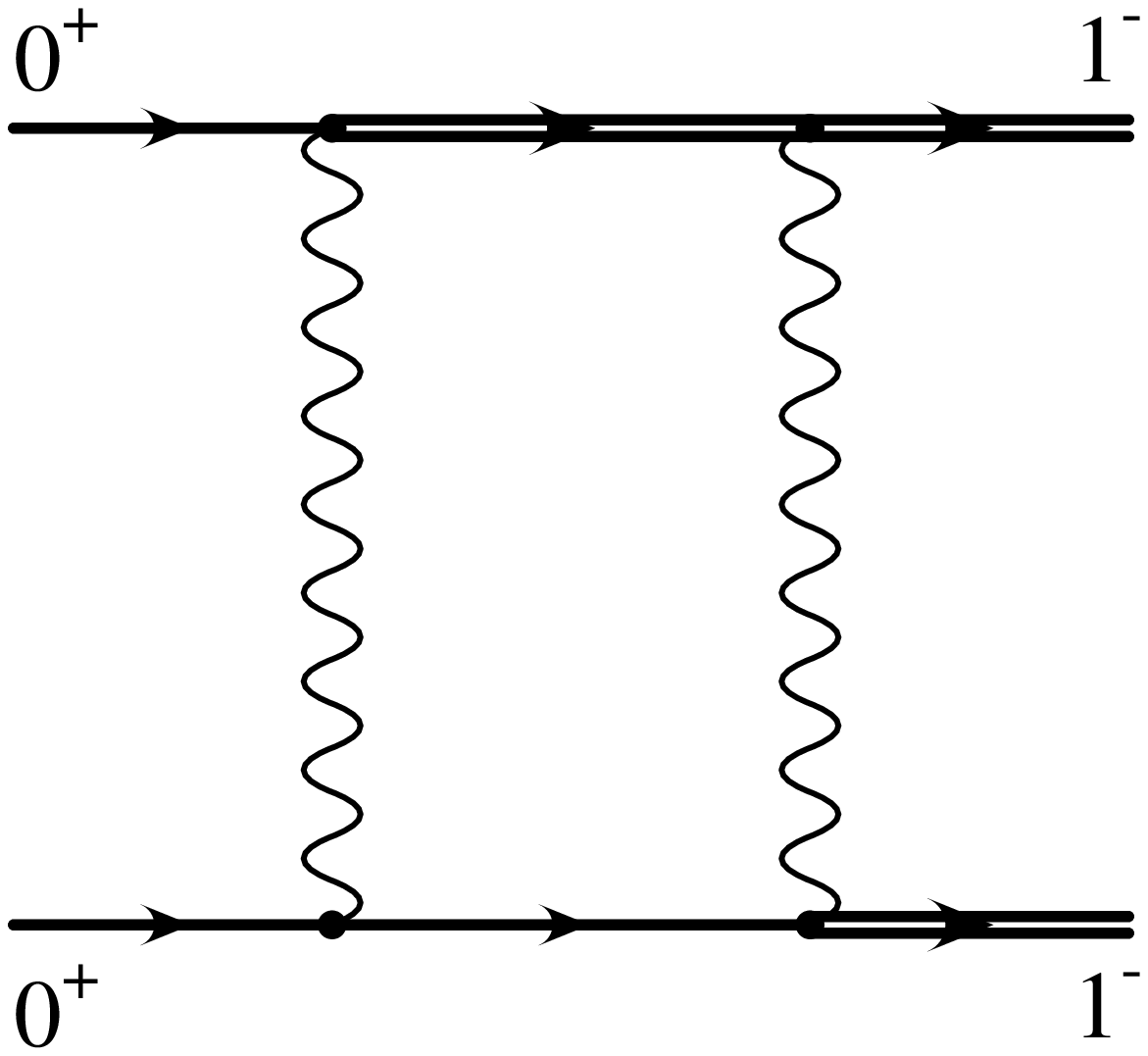}
\end{center}
\caption{Higher order process leading also to the excitation of both
ions to the GDR by the exchange of two photons. The contribution of
this process can be dominant over the one of
Fig.~\protect\ref{Fig:mutual} for large values of $Z_1$ and $Z_2$}
\label{Fig:higher}
\end{figure}

Of course the mutual excitation, as shown in Fig.~\ref{Fig:mutual},
is not the only process leading to two excited nuclei in the final
state.  There are also other ones, like the one shown in
Fig.~\ref{Fig:higher}, whose contribution can become more important
for higher $Z_1$ and $Z_2$ \cite{baur89}.

\section{Excitation to other states}
\label{Sec:other}

\begin{figure}[tbp]
\begin{center}
\ForceHeight{4cm}
\BoxedEPSF{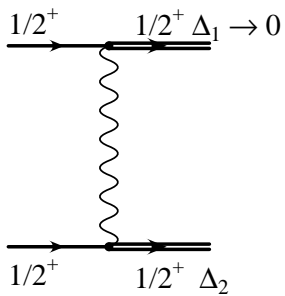}
\end{center}
\caption{Simultaneous excitation of two ions from a $1/2^+$ ground 
state to an excited $1/2^+$ excited state. In the limiting case of
$\Delta_1 \rightarrow 0$ we recover the usual formula of the
excitation due to the field of a magnetic dipole.}
\label{Fig:dipole}
\end{figure}

Let us look also at other multipolarities. Magnetic $M1$ transitions
can be treated in a similar way. As an example we look at the case of
a $1/2^+ \rightarrow 1/2^+$ excitation, as shown in
Fig.~\ref{Fig:dipole}.  As initial and final states have the same
quantum numbers, this case can also be used to study the limit of
elastic interaction, i.e., $\Delta_1 \rightarrow 0$. We expect to get
then the equivalent photon spectrum due to the static magnetic moment
of the nucleus. The equivalent photon spectrum for the static magnetic
moment was already studied in \cite{budnev75,baron93}. Here we are
looking at the excitation of a $1/2^+ \rightarrow 1/2^+$ transition in
the other nucleus as well.  The ion that emits the photon, has an
equivalent photon number
\begin{equation}
n_{\Delta_1}(\omega) = \frac{\alpha}{9 \pi} \frac{B(M1)}{R^2},
\qquad\qquad\mbox{(for $\omega < \gamma / R$)},
\end{equation}
where we have used the simple cutoff form factor and kept only the
leading term in the limit of $\gamma\rightarrow \infty$.  The
cross section for the simultaneous excitation is then given by
\begin{equation}
\sigma_{\Delta_1\Delta_2} = \frac{4 \pi \alpha^2}{81 R^2} 
B(M1,\Delta_1) B(M1,\Delta_2),
\end{equation}
which again is proportional to the product of the two $B(M1)$ values
and identical in form to Eq.~(\ref{Eq:approx}). The known result for
a static magnetic moment is obtained by replacing the $B(M1)$
value in the limit of $\Delta_1 \rightarrow 0$ by the magnetic moment
$\mu$ according to
\begin{equation}
B(M1) =  \frac{9}{4 \pi} \mu^2.
\end{equation}

On the other hand if we exchange the roles of both ions, we have the
situation where the equivalent photon spectrum of the $1/2^+
\rightarrow 1/2^+$ transition in the fast moving nucleus $A_2$ (with
$M1$ excitation energy $\Delta_2$) is absorbed by the static magnetic
moment of nucleus $A_1$ (with $\omega = \Delta_1 \rightarrow
0$). Although the integral over the equivalent photon spectrum taken
by itself diverges for $\omega \rightarrow 0$, the product with the
corresponding photoabsorption cross section (which vanishes for
$\omega \rightarrow 0$) leads to a constant limit.

It seems interesting to study this limiting case for other possible
transitions too, for example, for the transition $0^+ \rightarrow
0^+$.  This is a special case as the photon emission is dominated by
the $|M^C|^2$ matrix element with $J=0$ and no corresponding
$|T^e|^2$ exists. Therefore the photoexcitation cross section
vanishes for $J=0$ and we do not expect to be able to exchange to
role of both ions.

In principle the equivalent photon approach can also be extended to
other, higher multipolarities, for example, quadrupole
excitations. We may also study transitions with different
multipolarities for the two ions.  We have only looked briefly at
these processes, because the cross sections will be quite small. For
the $0^+ \rightarrow 2^+$ transition, we expect the cross section to
be proportional to $B(E2)/R^4$. These processes therefore become less
and less important, as the higher order processes, as shown in
Fig.~\ref{Fig:higher}, dominate the observed cross sections, see also
\cite{bertulani93}.

We now discuss some of the limitations of this approach, one has to be
aware of. One has to keep in mind that, in contrast to the elastic case, the
inelastic equivalent photon approach is more and more dominated by the
large values of $q_\perp$ when going to higher
multipolarities. Therefore the results are much more sensitive to the
detailed form of the form factors of both ions than in the elastic
case. The equivalent photon spectrum is therefore no longer a
property of the emitting nucleus alone, but of the absorbing nucleus
as well. This should become more and more important as the multipolarity is
increased. 

In the case of the photon-photon collisions,as studied in
\cite{hencken95}, these limitations are less important, because we
have two distinct transverse momentum scales: One scale is the limit
of the $q^2$ of the emitted photon.  This is given by the inverse of
the nuclear radius $R^{-1}$. The other one is the range of $q^2$,
where the virtual photon can be replaced a real photon. This scale is
normally given by the invariant mass of the produced system, which is
much larger than the nuclear scale. The only exceptions are the
$e^+$-$e^-$-production and to a smaller extent the
$\mu^+$-$\mu^-$-production.

\section{Acknowledgment}
\label{Sec:Ackno}

This work was partially supported by the Swiss National Science
Foundation (SNF) and the ``Freiwillige Akademische Gesellschaft''
(FAG) of the University of Basel. One of us (K.H.) would like to
thank them for their financial support. We would like to thank also 
G. Bertsch for the critical reading of the manuscript.

%
%

\end{document}